%% file: bjkxsec_sub_hepex.tex
\documentclass[aps,prd,preprint,floatfix,nofootinbib,superscriptaddress,showpacs]{revtex4}
\usepackage{graphicx}

 \input{prd_definitions}
 
\begin{document}

 \bibliographystyle{apsrev}

 \title {Measurement of the  {\boldmath $B^+$} production cross section
         in {\boldmath $p \bar{p}$} collisions at {\boldmath $\sqrt{s}=$}
         1960 GeV}
\input{Oct_2006_Authors_alt_removed}
\noaffiliation

 \begin{abstract}
   We present a new measurement of the $B^+$ meson differential cross
 section $d \sigma/d p_T$ at $\sqrt{s}=1960$ GeV.
 The data correspond to an integrated luminosity of 739 pb$^{-1}$
 collected with the upgraded CDF detector (CDF II) at the Fermilab Tevatron collider.
  $B^+$  candidates are reconstructed through the decay $B^+ \rightarrow J/\psi\; K^+$,
 with $ J/\psi \rightarrow \mu^+ \; \mu^-$. The integrated cross section
  for producing $B^+$ mesons with $p_T \geq 6 \;\gevc$ and  $|y| \leq 1$
  is measured to be $2.78 \pm 0.24\;\mu$b.
 \end{abstract} 
 \pacs{13.85.Ni, 12.38.Qk, 13.25.Hw, 14.40.Nd }
 \preprint{FERMILAB-PUB-06-458-E}  
 \maketitle
 \section {Introduction}  \label{sec:ss-intro}
  Measurements of the bottom quark production cross section at the Tevatron
  collider
  probe the ability of perturbative QCD to predict absolute rates in
  hadronic collisions. At the perturbative level,
  calculations of the hard scattering cross sections have been carried out
  at next-to-leading order (NLO)~\cite{nlo} and also
  implemented with logarithmic $p_T^b/m_b$~\footnote{Mass ($m_b$) and
  transverse momentum ($p_T^b$)
  of the bottom quarks involved in the hard scattering.}
  corrections  evaluated  to next-to-leading logarithmic
  accuracy (NLL)~\cite{nll}. In both cases, these QCD predictions are affected by
  large theoretical uncertainties such as the dependence on the choice
  of the renormalization
  and factorization scales and the $b$-quark mass~\cite{depen,cacc}.
  Accurate measurements 
  could help in improving the theoretical prediction.
  Unfortunately, as noted in 
  Ref.~\cite{bstatus}, measurements of the $b$-quark cross section at the Tevatron 
  appear to be inconsistent among themselves. 
  Reference~\cite{bstatus} uses
  the prediction of a NLO
  calculation~\cite{nlo} implemented with a non-perturbative model for the $b$-quark
  fragmentation~\footnote{
 This calculation uses a $b$-quark mass of
 $m_b=4.75 \; \gevcc$, renormalization and factorization scales
 $\mu_R=\mu_F=\sqrt{p_T^2 +m_b^2}$, the MRSD$_0$~\cite{mrsd0}
 fit to the parton distribution functions (PDF), and a fragmentation
 fraction $f_u=0.375$. The fragmentation model
 is based on the Peterson fragmentation function~\cite{pet} with the $\epsilon$ parameter
 set to 0.006 according to fits to $e^+e^-$ data~\cite{chrin}.} in order to
 compare all measurements performed at the Tevatron. This calculation
  predicts
   $\sigma(p_T^{B^+} \geq 6 \; \gevc, |y|^{B^+} \leq 1)=  0.9  \;\mu$b.
 Previous  measurements~\cite{cdfb1,cdfb2} performed by the CDF collaboration
 at $\sqrt{s}=1.8 $ TeV yield
 $\sigma(p_T^{B^+} \geq 6 \; \gevc, |y|^{B^+} \leq 1) = 2.66 \pm 0.61$ 
  and  $3.6 \pm 0.6$  $\mu$b, respectively.
  The ratios of these measurements to the NLO prediction are
  $(2.9 \pm 0.67) $ and  $(4.0 \pm 0.6)$, respectively.
  In contrast, the ratios of the CDF and D${\not\!{\rm O}}$ measurements of the
  $b$ cross section, that are not based upon the detection 
  of $J/\psi$ mesons~\cite{cdfb3,cdfb4,d0b1,d0b2,d0b3},
  to the same theoretical prediction have an appreciably smaller
  average (2.2 with a 0.2 RMS deviation~\cite{bstatus}).
 The cause of the inconsistency could be  experimental difficulties inherent to each
 result or some underlying, and not yet appreciated, production of new
 physics.  Therefore, it is of interest
 to clarify the experimental situation.

 This paper presents a new measurement of the $B^+$ production cross section
 that uses fully reconstructed $B^\pm \rightarrow J/\psi \; K^\pm$ decays.
 We  follow closely the experimental procedure used in
 Refs.~\cite{cdfb1,cdfb2}, but we simplify the analysis selection criteria 
 in order to reduce systematic uncertainties. 
 The $B^+$ production cross section  is the ratio of the number
 of observed  $B^+$ candidates to the product of the detector acceptance,
 integrated luminosity, and branching fraction of the decay
 $B^+\rightarrow J/\psi \; K^+$ with $  J/\psi\rightarrow \mu^+\; \mu^-$.  
 Section~\ref{sec:det} describes the detector systems relevant to this analysis.
 The data collection, event selection, and $B^\pm$  reconstruction are described in
 Sec.~\ref{sec:anal}. In
 Sec.~\ref{sec:diffsec}, we evaluate the detector acceptance
 and derive the total and differential
 $B^+$ cross section. Our conclusions are presented in Sec.~\ref{sec:concl}.
 \section{The CDF~II detector}
 \label{sec:det}
 CDF is a multipurpose detector, equipped with a charged particle spectrometer
 and a finely segmented calorimeter. In this section, we describe the detector components
 that are relevant
 to this analysis. The description of these subsystems can be found
 in Refs.~\cite{det1,det2,det3,det4,det5,det6,det7}. Two devices inside the 1.4 T solenoid
 are used for measuring the momentum of charged particles: the silicon vertex detector (SVX II) and
 the central tracking chamber (COT). The SVX~II consists of double-sided microstrip sensors
 arranged in five cylindrical shells with radii between 2.5 and 10.6 cm. The detector is divided
  into three contiguous five-layer sections along the beam direction for a total $z$ coverage~\footnote{
 In the CDF coordinate system, $\theta$ and $\phi$ are the polar and azimuthal angles
 of a track, respectively,
 defined with respect to the proton beam direction, $z$. The pseudorapidity $\eta$
 is defined as $-\log \;\tan (\theta/2) $.
 The transverse momentum of a particle is $p_T=P \; \sin (\theta)$.
 The rapidity is defined as $y=1/2 \cdot \log ( (E+p_z)/(E-p_z) )$, where $E$ and $p_z$ are
 the energy and longitudinal momentum of the particle associated with the track.}
  of 90 cm.
 The COT is a cylindrical drift chamber containing 96 sense wire layers grouped
 into eight alternating superlayers of axial and stereo wires.
 Its active volume covers $|z| \leq 155$ cm and 40 to 140 cm in radius.
 The central muon detector (CMU) is located around the central electromagnetic and
 hadronic calorimeters, which have a thickness of 5.5 interaction lengths
 at normal incidence.

 The CMU detector covers a nominal pseudorapidity range $|\eta| \leq 0.63$ relative to the
 center of the detector, and  is segmented into two barrels of 24 modules, each covering
 15$^\circ$ in $\phi$. Every module is further segmented  into three  submodules, each
 covering 4.2$^\circ$ in $\phi$ and consisting of four layers of drift chambers.
   The smallest drift unit, called a stack, covers a 1.2$^\circ$ angle in $\phi$.
 Adjacent pairs of stacks are combined together into a tower. A track segment
(hits in two out of four layers of a stack) detected
in a tower is referred to as a CMU stub.
 A second set of muon drift chambers (CMP) is located behind an additional steel absorber
 of 3.3 interaction lengths.  Muons which produce a stub in both CMU and CMP systems
 are called CMUP muons.

 The luminosity is measured using gaseous Cherenkov counters (CLC) that monitor
 the rate of inelastic $p\bar{p}$ collisions. The inelastic $p\bar{p}$ cross section at
 $\sqrt{s}=1960$ GeV is scaled from measurements at  $\sqrt{s}=1800$ GeV using
 the calculations in Ref.~\cite{sigmatot}. The integrated luminosity is 
 determined with a 6\% systematic accuracy~\cite{klimen}.
 
 CDF uses a three-level trigger system. At Level 1 (L1), data from every beam crossing are stored
 in a pipeline capable of buffering data from 42 beam crossings.
 The L1 trigger either rejects events or copies them into one of the four Level 2 (L2)
 buffers. Events that pass the L1 and L2 selection criteria
 are sent to the Level 3 (L3) trigger, a cluster of computers running  speed-optimized
 reconstruction code.  

 For this study, we select events with two muon candidates identified by the L1
 and L2 triggers. The L1  trigger uses tracks with $p_T \geq 1.5 \; \gevc$ found by
 a fast track processor (XFT). The XFT examines COT hits from four axial superlayers and
  provides $r-\phi$ information.
 The XFT finds tracks with  $p_T \geq 1.5 \; \gevc$ in azimuthal sections of 1.25$^\circ$.
 The XFT passes the tracks to a set of extrapolation units that determines the CMU
 towers in which a CMU stub  should be found if the track is a muon.
 If a stub is found, a L1 CMU primitive is generated.
 The L1 dimuon trigger  requires at least two CMU primitives, separated by at least two CMU towers.
  At L1, there is no requirement that muons have opposite
 charge. During the data-taking period in which the dimuon sample used for this
 analysis was collected, the Tevatron luminosity has increased from
 1 to 100$\times 10^{30}$~cm$^{-2}$~s$^{-1}$. Accordingly, the L2 trigger, that
 started with no additional requirement, has incrementally required dimuons
 with opposite charge, opening azimuthal angle $\delta \phi \leq 120^\circ$, and
 $p_T \geq 2 \; \gevc$. All these trigger requirements are mimicked by the detector
 simulation on a run-by-run basis.
 At L3, muons are required to have opposite charge,  invariant mass in the window
 $2.7-4.0 \; \gevcc$, and $|\delta z_0| \leq 5$ cm, where $z_0$ is the $z$ coordinate of the muon track
 at its point
 of closest approach to the beam line in the $r-\phi$ plane.
 These requirements define the $J/\psi\rightarrow \mu^+ \mu^-$ trigger.

 We use two additional triggers in order to verify 
 the detector simulation.
 The first trigger (CMUP$p_T$4) selects events with at least 
 one L1 and one L2 CMUP primitive with $p_T \geq 4 \; \gevc$, and an additional
 muon found by the L3 algorithms. Events collected with this
 trigger are used to measure the muon trigger efficiency.
 The second trigger ($\mu-$SVT) requires a L1 CMUP primitive
 with $p_T \geq 4 \; \gevc$ accompanied by a L2 requirement
 of an additional XFT track with  $p_T \geq 2 \; \gevc$
 and displaced from the interaction point. These events are used
 to verify the muon detector acceptance and the muon reconstruction
 efficiency.
  \section{Data selection and {\boldmath  $B^\pm$} reconstruction}
 \label{sec:anal}
 We search for $B^\pm \rightarrow J/\psi K^\pm$ candidates in the data set
 selected by the $J/\psi\rightarrow \mu^+ \mu^-$  trigger.
 Events are reconstructed off-line taking advantage of more refined calibration
 constants and reconstruction algorithms.

 The transverse momentum resolution of tracks reconstructed using COT hits
 is $\sigma(p_T)/p_T^2 \simeq 0.0017\; [\gevc]^{-1}$. COT tracks are extrapolated
 into the SVX~II detector and refitted adding hits consistent with the track
 extrapolation.
 Stubs reconstructed in the CMU detector are matched to tracks with $p_T \geq 1.3 \; \gevc$.
 A track  is identified as a CMU muon if $\Delta r\phi$,
 the distance in the $r-\phi$ plane between the track
 projected to the CMU chambers and a CMU stub, is less than 30 cm.
 We also require that muon-candidate stubs correspond to a L1 CMU primitive,
  and correct the muon momentum
 for energy losses in the detector.

 We search for $J/\psi$ candidates by using pairs of CMU muons with opposite charge,
 and $p_T \geq 2 \; \gevc$ (this  requirement avoids the region of rapidly changing
  efficiency around the trigger threshold).
 The invariant mass of a muon pair is evaluated by
 constraining the two muon tracks to originate from a common point in
 three-dimensional space
 (vertex constraint) in order to improve the mass resolution.
 All muon pairs with invariant mass  in the range $3.05-3.15\; \gevcc$ 
 are considered to be $J/\psi$ candidates.
 
 If a $J/\psi$ candidate is found, we search for $B^\pm$ mesons by considering 
 all remaining charged particle tracks in the 
 event as possible kaon candidates. As in  previous measurements~\cite{cdfb1,cdfb2},
 we select tracks with $p_T \geq 1.25 \; \gevc$ and with $|\delta z_0| \leq 1.5$ cm
 with respect to the $z_0$ position of the  $J/\psi$ candidate. We require that
 kaon-candidate tracks
 have at least 10 hits in both COT axial and stereo superlayers. This limits
 the pseudorapidity acceptance to $|\eta| \leq 1.3$.
 The invariant mass of the  $\mu^+\; \mu^- \; K^\pm$ system
 is evaluated  constraining the corresponding tracks to have a common origin
 while the $\mu^+\; \mu^-$ invariant mass is constrained to the value of
 3.0969 $\gevcc$~\cite{pdg}. As in  Refs.~\cite{cdfb1,cdfb2}, we select
 $B^\pm$ candidates with $p_T \geq  6 \; \gevc$. From the pseudorapidity
 acceptance of CMU muons ($|\eta| \leq 0.8$) 
  and  the $p_T$ cuts on the
 $\mu^\pm$ and $B^\pm$ transverse momenta, it follows that: (1) no kaon from $B^\pm$ decays
 is emitted at $|\eta| \geq 1.3$; (2)  the reconstructed
 $B^\pm$ candidates have rapidity $|y| \leq1$.

\begin{figure}
 \begin{center}
 \leavevmode
 \includegraphics*[width=\textwidth]{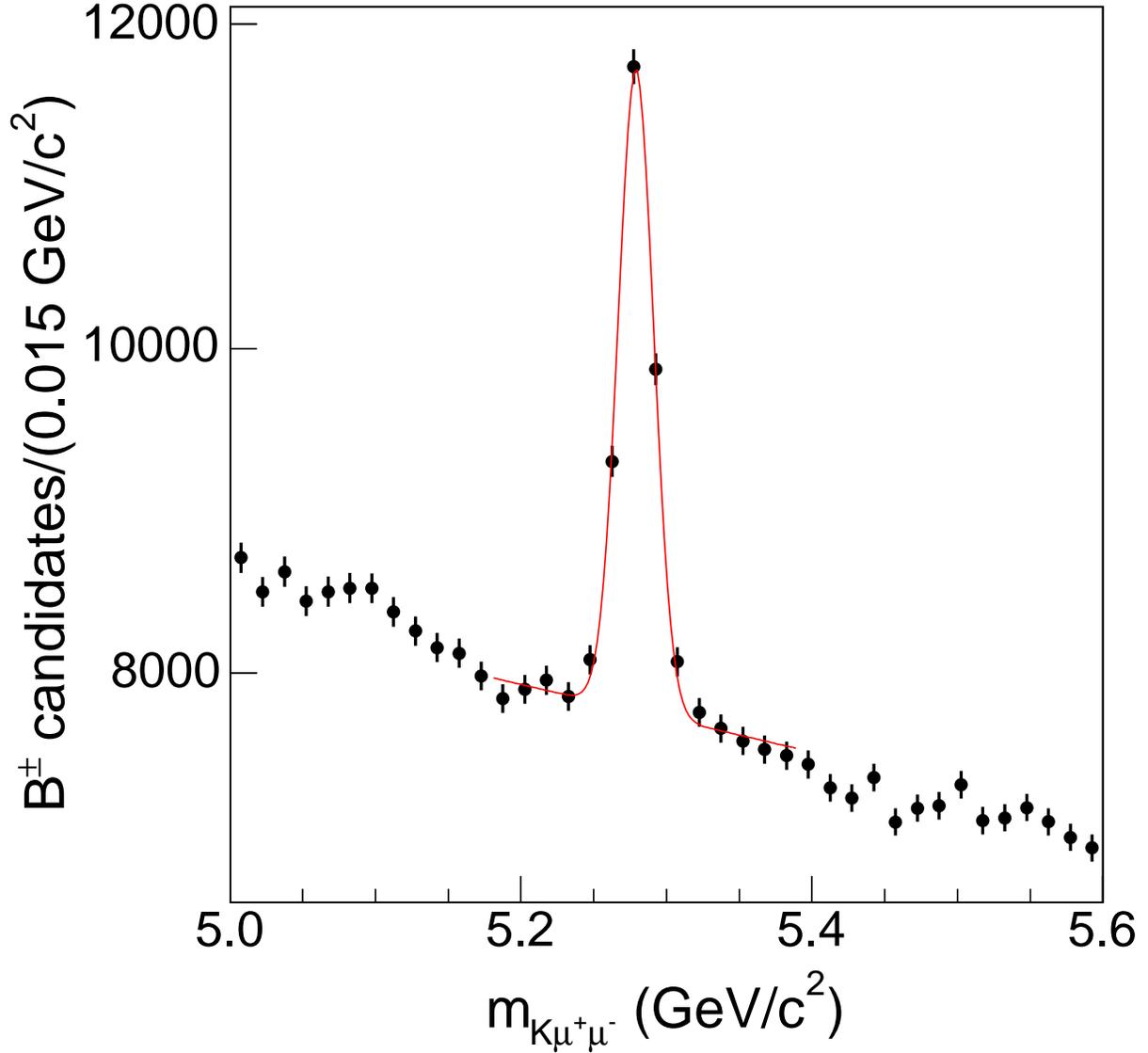}
 \caption[]{Invariant mass distribution of all $B^\pm$ candidates. The line
 represents a fit to the data using a first order polynomial plus a Gaussian
 function in order to estimate the background and the $B^\pm$ signal, respectively.} 
 \label{fig:fig1}
 \end{center}
 \end{figure}

 In contrast with the analyses in  Refs.~\cite{cdfb1,cdfb2},
 we do not require 
  the proper decay length of the  $B^\pm$ candidates to  be larger than 100 $\mu$m.
 By doing so, we avoid  two large sources of systematic uncertainty:
 (1) the simulated efficiency of the  SVX~II detector; (2) the dependence
 of the decay length distribution on the
 simulated SVX~II resolution and $B^\pm$ transverse momentum distribution.
 The invariant mass distribution of all $B^\pm$ candidates found in this study is shown
 in Fig.~\ref{fig:fig1}.
  \section{Differential cross section}
 \label{sec:diffsec}
 
 To measure the differential cross section,
 we divide the sample of $B^\pm$ candidates into five  $p_T$ bins: $6-9$,
 $9-12$, $12-15$, $15-25$, and $\geq 25 \; \gevc$~\cite{cdfb1,cdfb2}.
 In each  $p_T$ bin, we fit the invariant mass distribution
 of the $B^\pm$ candidates  with a
 binned maximum-likelihood method to determine the number 
 of $B^\pm$ mesons. The fit likelihood uses
 a Gaussian function to model the $B^\pm$ signal.
 As in the previous measurements~\cite{cdfb1,cdfb2},
 we use a first order polynomial to estimate the
 underlying combinatorial background. 
  As shown by Fig.~\ref{fig:fig1},
 the mass region above the $B^\pm$ meson signal
 is not affected by partially reconstructed $B^\pm$ decays and
  is quite well described by a straight line.
  We fit the data in the invariant mass
  range $5.18-5.39 \;\gevcc$. The lower limit is chosen to avoid the region populated by
  partially reconstructed $B^\pm$ decays. The width of the fitted mass range
  determines the statistical error of the background estimate.
  Since we have a much larger data set than previous measurements~\cite{cdfb1,cdfb2}, we
  can afford to fit the data in a smaller mass range  in order to reduce
  the systematic uncertainty due to the background modeling. 
 
\begin{figure}
 \begin{center}
 \leavevmode
 \includegraphics*[width=\textwidth]{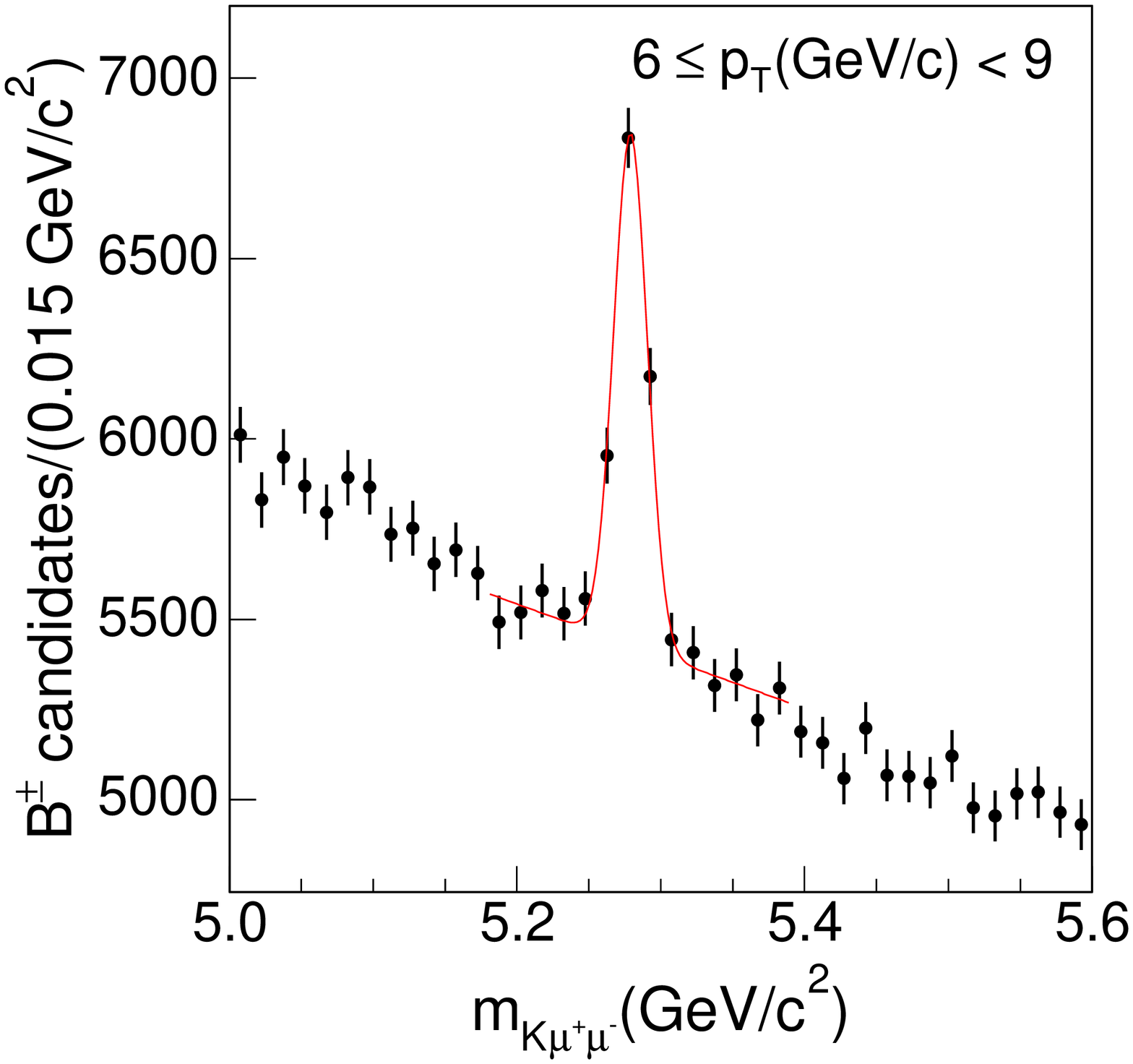}
 \caption[]{Invariant mass distribution of $B^\pm$ candidates with
 $6\leq p_T \leq 9 \; \gevc$. The line represents the best fit to the data described in the text.} 
 \label{fig:fig2}
 \end{center}
 \end{figure}

\begin{figure}
 \begin{center}
 \leavevmode
 \includegraphics*[width=\textwidth]{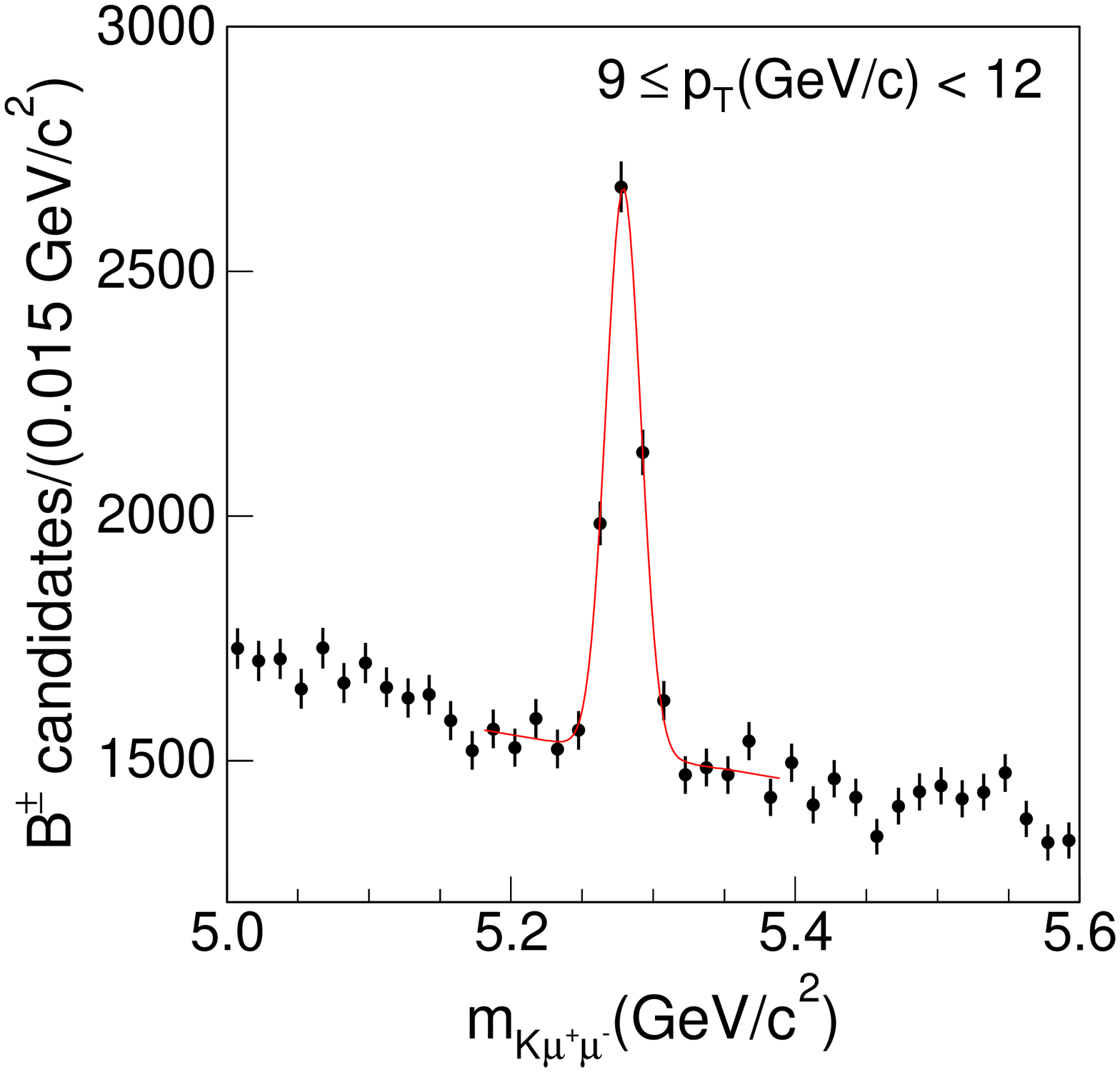}
 \caption[]{Invariant mass distribution of $B^\pm$ candidates with
 $9\leq p_T \leq 12 \; \gevc$. The line represents the best fit to the data described in the text.} 
 \label{fig:fig3}
 \end{center}
 \end{figure}

\begin{figure}
 \begin{center}
 \leavevmode
 \includegraphics*[width=\textwidth]{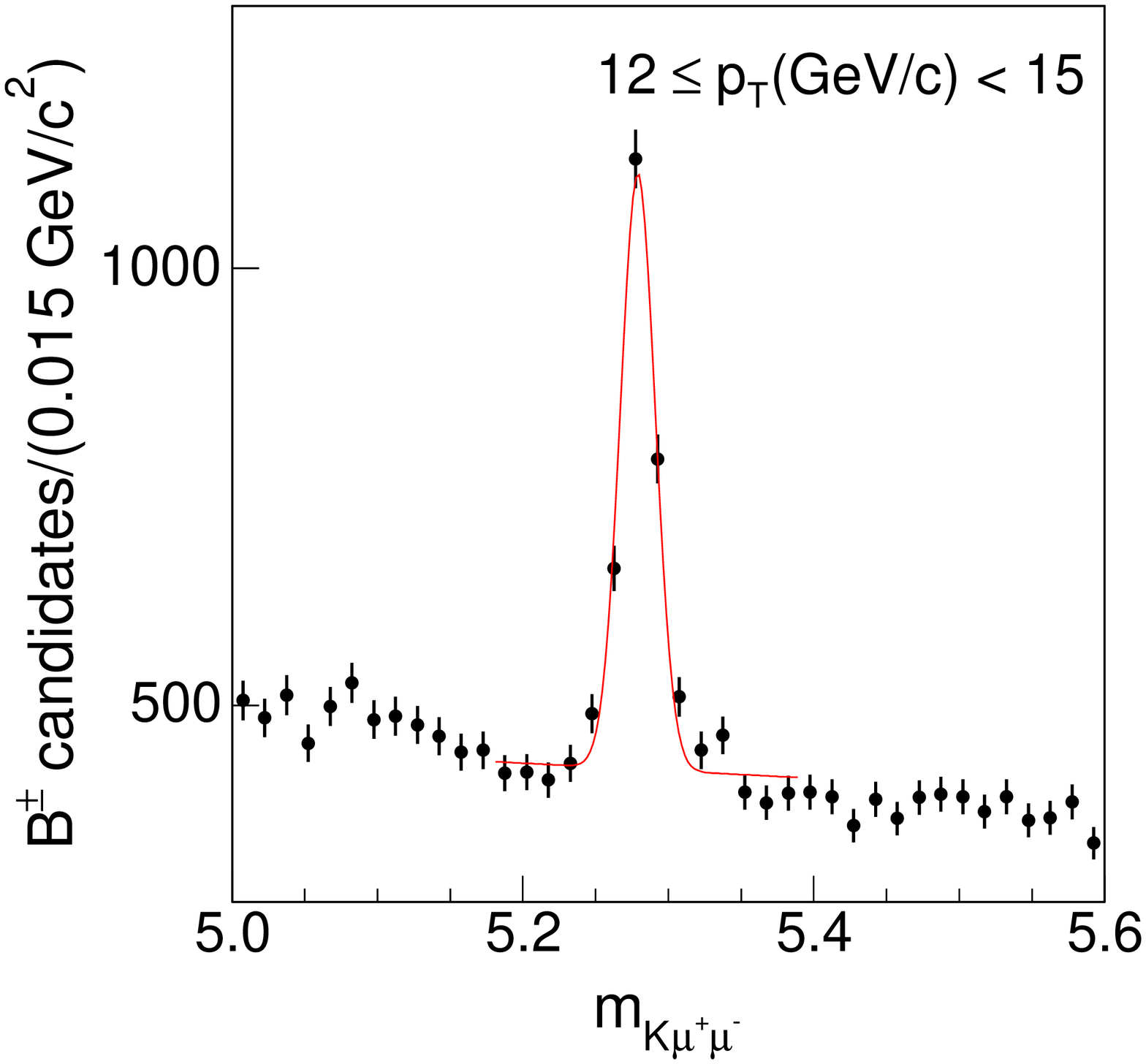}
 \caption[]{Invariant mass distribution of $B^\pm$ candidates with
 $12\leq p_T \leq 15 \; \gevc$. The line represents the best fit to the data described in the text.} 
 \label{fig:fig4}
 \end{center}
 \end{figure}

\begin{figure}
 \begin{center}
 \leavevmode
 \includegraphics*[width=\textwidth]{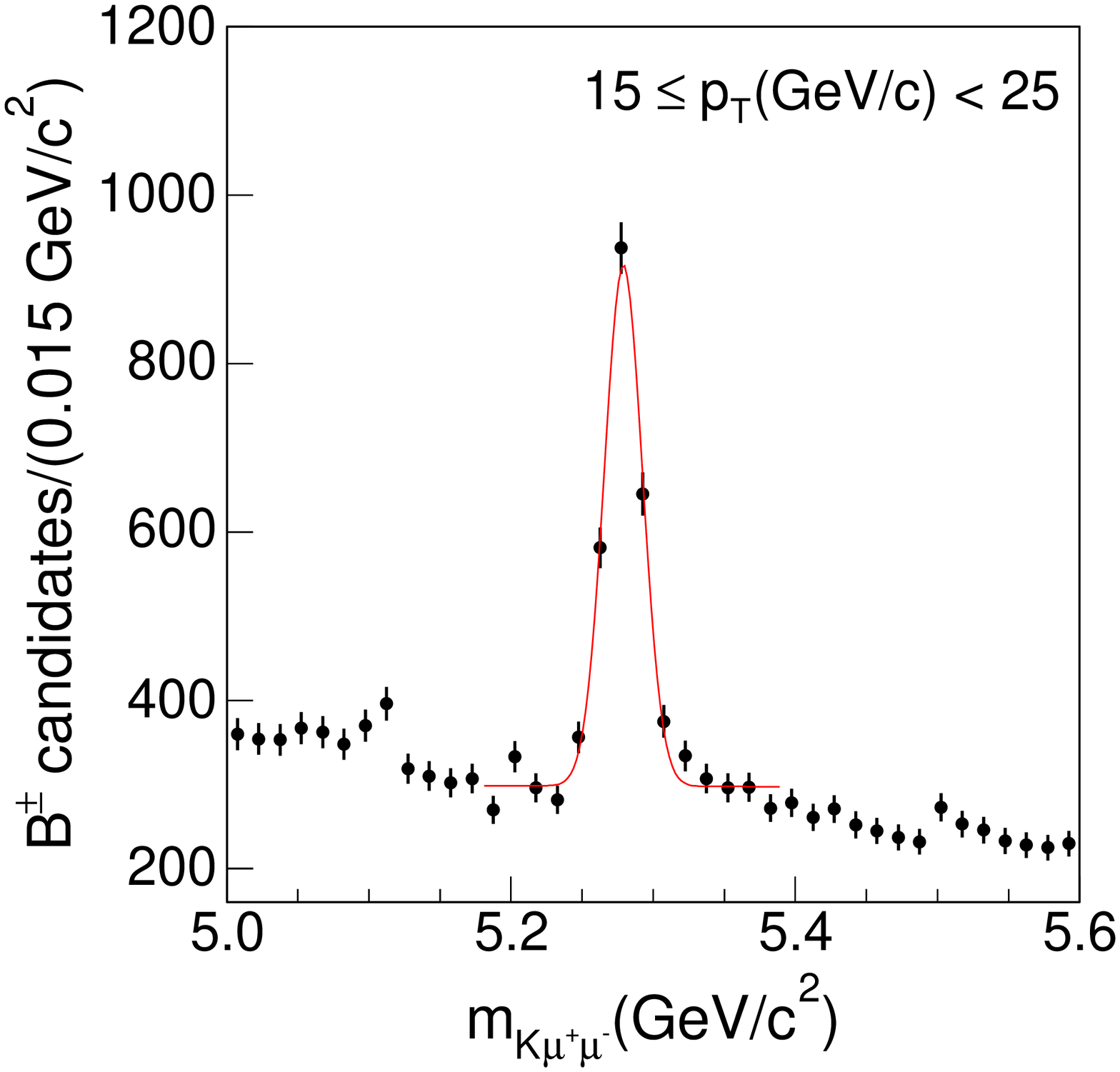}
 \caption[]{Invariant mass distribution of $B^\pm$ candidates with
 $15\leq p_T \leq 25 \; \gevc$. The line represents the best fit to the data described in the text.} 
 \label{fig:fig5}
 \end{center}
 \end{figure}

\begin{figure}
 \begin{center}
 \leavevmode
 \includegraphics*[width=\textwidth]{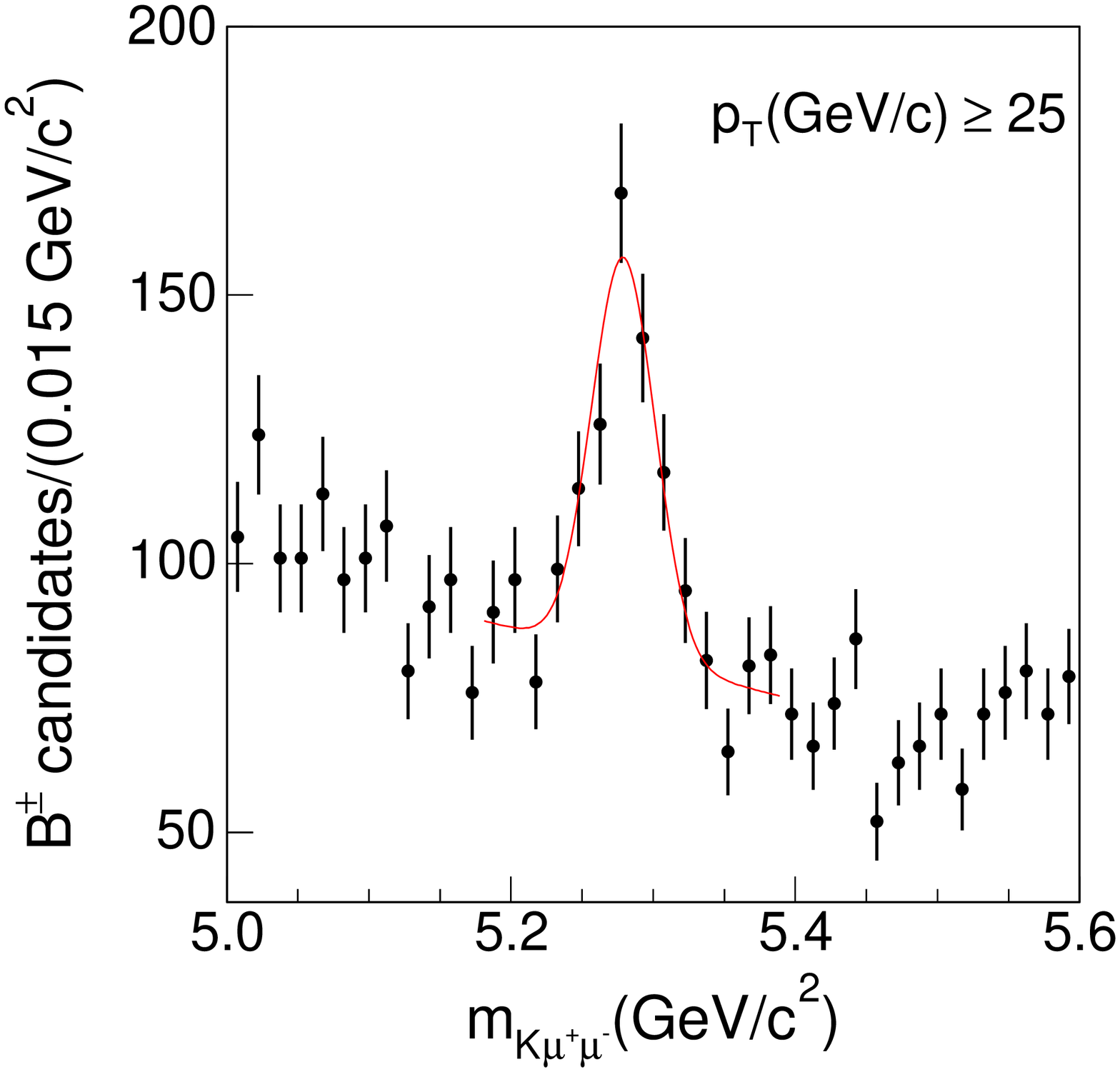}
 \caption[]{Invariant mass distribution of $B^\pm$ candidates with
 $ p_T \geq 25 \; \gevc$. The line represents the best fit to the data described in the text.} 
 \label{fig:fig6}
 \end{center}
 \end{figure}

 The average  of the  $B^\pm$ mass values returned by the fits
 in the different  $p_T$ bins is $5.2790 \; \gevcc$ with
 a  0.5 MeV$/c^2$ RMS deviation,  in agreement with the
 PDG value~\cite{pdg}. In the fit used to determine the number of
  $B^\pm$ mesons, we fix the $B^\pm$ mass value to $5.279 \; \gevcc$~\cite{pdg}.
  The width of the Gaussian is a free fit parameter; 
   the value of  $\sigma$ returned by the fit increases from $12.0\pm0.4$ 
  to $20.0\pm0.4$  MeV/$c^2$ from the first to last $p_T$ bin, in agreement with the
 simulation prediction.
  The fits are shown in
 Figs.~\ref{fig:fig2} to~\ref{fig:fig6}. They
 return a signal of $2792 \pm 186$,
 $2373 \pm 110$, $1365 \pm 66$, $1390\pm 63$,  and $277 \pm 44$
 $B^\pm$ mesons in the five $p_T$ bins.

 We have investigated possible systematic uncertainties in the fit results.
 We have studied the contribution
 of the  $B \rightarrow J/\psi\; \pi$ decay mode, the branching
 fraction of which is ($4.9\pm0.6$)\%  of that of
 the  $B \rightarrow J/\psi\; K$ decay mode~\cite{pdg}.  
 As shown in Ref.~\cite{kcorr}, the invariant mass distribution of 
 Cabibbo-suppressed $B$ decays,  reconstructed assuming that pions are kaons,
 is shifted  into the mass region $5.28 - 
 5.44 \; \gevcc$, which partially overlaps with that of the  $B \rightarrow J/\psi \; K$
 decay mode. However, part of the Cabibbo-suppressed contribution
 is also used by the fit to predict the  background under the
 $B \rightarrow J/\psi \; K$ signal with the effect of
 reducing its size.
 When adding  the expected contribution
 of the  Cabibbo-suppressed decays, the
  $B \rightarrow J/\psi \; K$ signal returned by the fit
 decreases by $(1\pm1)$\%.
  We have investigated other possible causes of systematic uncertainties
  in the $B$ signal estimate.
  We have compared the results of our fit with those returned using
  an  unbinned likelihood method.
  We have
  decreased the fitted mass range to $5.24-5.33 \;\gevcc$, and
  we have fitted the larger mass interval   $5.18-5.60 \;\gevcc$.
  We have fitted the signal with two
  Gaussian functions in order to study
 detector resolution effects. The $B$ signal returned by these fits
 does not vary by more than $\pm 1.5$\%.
 Therefore, we attribute an overall $\pm$2\% systematic uncertainty
 to the fit results.   
 \begin{table}
 \caption{Detector acceptance, $\cal A$,  as a function of the $B^\pm$ $p_T$.
 The acceptance $\cal A_{\rm corr}$ includes corrections evaluated using the data.
 The average $<p_T>$ is the value at which the theoretical differential cross
 section~\cite{nlo} equals the integrated cross section in each momentum bin divided
 by the bin width.} 
\begin{center}
\begin{ruledtabular}
\begin{tabular}{cccc}
  $p_T$ range ($\gevc$)& $<p_T>$  ($\gevc$)  & $\cal A$ (\%) & $\cal A_{\rm corr}$ (\%) \\
   $6-9$ & 7.37    & 1.545 & $1.780 \pm 0.045$ \\
   $9-12$ & 10.38   & 3.824 & $4.405 \pm 0.111$ \\
    $12-15$ & 13.39 & 5.966 & $6.872 \pm 0.173$  \\
    $15-25$ & 19.10 & 8.819 & $10.16 \pm 0.25$ \\
    $\geq 25$ &     & 12.516 & $14.42 \pm 0.36$ \\
\end{tabular}
 \end{ruledtabular}
 \end{center}
 \label{tab:tab1}
 \end{table}
\subsection{Acceptances and Efficiencies}
\label{sec:accor}
 The detector acceptance is calculated with a Monte Carlo simulation
 based upon a  NLO calculation~$^2$. The  $B^\pm$ decay is modeled with the
 {\sc evtgen} Monte Carlo program~\cite{evtgen} that accounts for the
 $J/\psi$ longitudinal polarization~\cite{pol}.
 The detector response to particles produced by $B^\pm$ decays  is 
 modeled with the CDF~II detector simulation that in turn
  is based on the {\sc geant}  Monte Carlo program~\cite{geant}.
  The simulation includes the generation of L1 CMU trigger primitives.
  Simulated events are processed and selected with the same
 analysis code used for the data.
 The acceptances estimated using the simulation are listed in Table~\ref{tab:tab1}.
 We use the data to verify  the detector acceptance and efficiencies
  evaluated  using the CDF~II detector
 simulation. We adjust the simulation to match measurements in the data of: (1) the off-line
 COT track reconstruction efficiency; (2) the CMU detector acceptance and
 efficiency; (3) the efficiency for finding L1 CMU primitives;
  and (4) the efficiency
  of the L1, L2, and L3 triggers. 

 In the simulation, the  off-line
 COT track reconstruction efficiency is given by the fraction of tracks, which
 at generator level satisfy the  $p_T$ and $\eta$ selection cuts,  that
 survive after selecting fully simulated events as the data.
 The COT track reconstruction efficiency is found to be $0.998 \pm 0.002$. 
 The  same efficiency in the data is measured by embedding  COT hits generated from 
 simulated tracks into $J/\psi$ data. In Ref.~\cite{bishai},
  the
 COT track reconstruction efficiency  in the data
 is measured to be 0.996 with a $\simeq 0.006$ systematic
 accuracy~\footnote{The efficiency measurement
 was performed in a subset of the data used for this analysis.
 Studies of independent data samples collected in the data taking period
  used for this analysis show that changes of the track reconstruction efficiency 
  are appreciably smaller than the quoted systematic uncertainty~\cite{matt}.}.
 We conclude that the efficiencies
  for reconstructing the $ \mu^+\; \mu^- K^\pm$
 system in the data and the simulation are equal within a 2\% systematic error.
 Kaon decay and interactions are modeled with the CDF~II detector simulation. 
 Because of the uncertainties of the detector materials and the
 nuclear interaction cross sections, the kaon 
 tracking efficiency has an additional 0.3\% uncertainty~\cite{giagu}.

\begin{figure}
 \begin{center}
 \leavevmode
\includegraphics*[width=\textwidth]{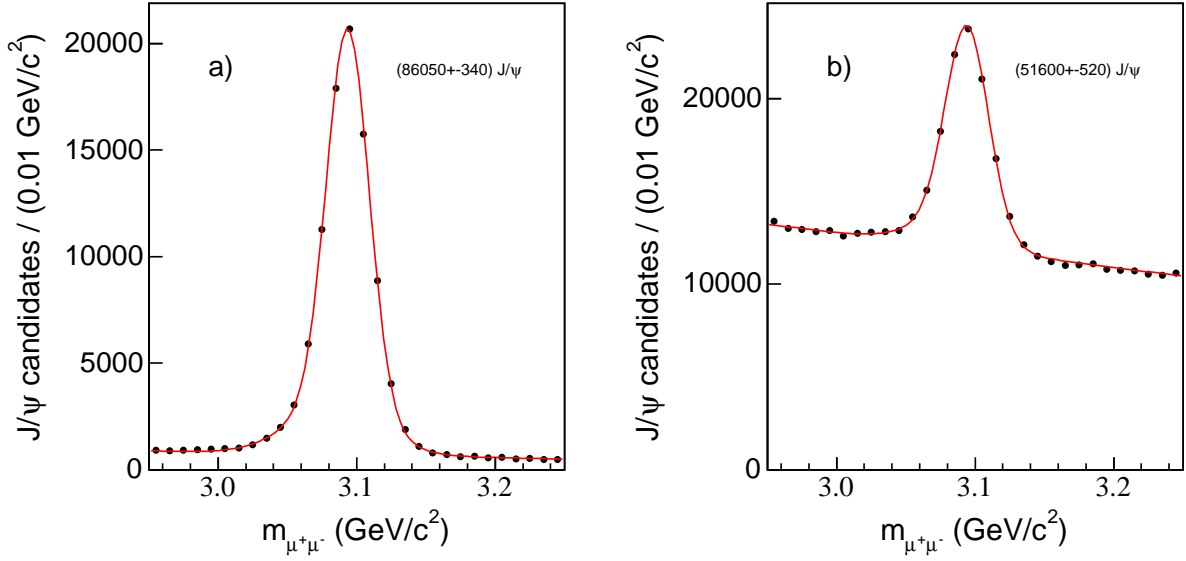}
 \caption[]{Invariant mass distribution of a CMUP muon paired with all
 charged tracks in the event
   with (a) or without (b) a CMU stub.}
 \label{fig:fig7}
 \end{center}
 \end{figure}

 In the simulation, the fraction of CMU stubs generated by muon tracks with
 $p_T \geq 2\; \gevc$ and $|\eta| \leq 0.8$ is $0.6439 \pm 0.0004$. In the data, this
 efficiency is measured by using  $J/\psi \rightarrow \mu^+\;\mu^-$ decays
 acquired with the $\mu$-SVT trigger. 
   We evaluate the invariant mass of 
  all pairs of a CMUP track and a track with displaced impact parameter,
  $p_T\geq 2 \; \gevc$, and $|\eta| \leq 0.8$.
  We fit the invariant mass distribution
  with a first order polynomial plus two
  Gaussian functions to extract the $J/\psi$ signal.
 From the number
 of $J/\psi$ mesons reconstructed using displaced tracks with or without
 a CMU stub (Fig.~\ref{fig:fig7}(a) and~(b), respectively),
  we derive an efficiency of $0.6251 \pm  0.0047$.
 The integrated  efficiency is evaluated after having weighted the $p_T$ and $\eta$ distributions of 
 displaced tracks in the data to be equal to those
 of muons from $B$ decays in the simulation.

\begin{figure}
 \begin{center}
 \leavevmode
 \includegraphics*[width=\textwidth]{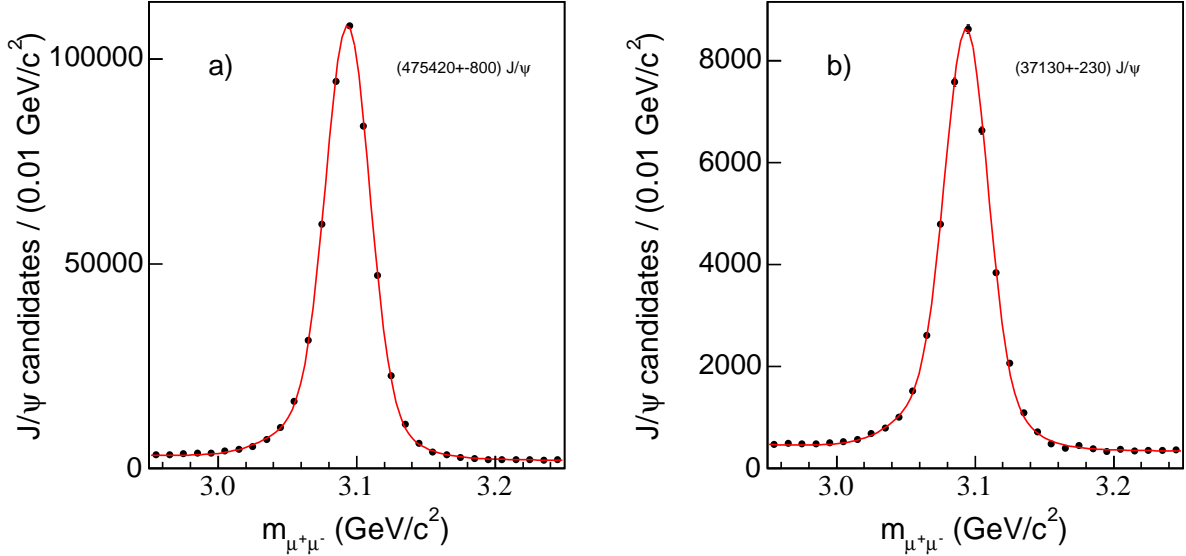}
 \caption[]{Invariant mass distribution of a CMUP muon paired with all
 CMU muons in the event
   with (a) or without (b) a L1 CMU primitive.}
 \label{fig:fig8}
 \end{center}
 \end{figure}

  In the simulation, the efficiency for finding a CMU primitive (CMU stub matched by a XFT track)
  is $0.8369 \pm 0.0004$. This efficiency is measured in the data by using events acquired with the
   CMUP$p_T$4 trigger.
  We combine the CMUP muon with all other CMU muons
  found in the event with and without a L1 CMU primitive.
  We extract the number of
  $J/\psi \rightarrow \mu^+ \mu^-$ mesons  by fitting the invariant mass distributions
  of all candidates with a first  order polynomial plus two Gaussian functions.
  By comparing the fitted numbers of   $J/\psi$ candidates with and without
  L1 CMU primitive (Fig.~\ref{fig:fig8}(a) and~(b), respectively)
  we derive an efficiency of $0.9276 \pm 0.0005$.
 The integrated efficiency is evaluated after having
  weighted the $p_T$ and $\eta$ distributions of the additional CMU muons to be equal to that
 of muons from $B$ decays in the simulation.

 In the simulation, the efficiencies of the L1 and L2 triggers are
  0.9868 and 0.9939, respectively.
 By studying  $J/\psi$ candidates acquired with the  CMUP$p_T$4
 trigger,  the L1 efficiency is measured to be $0.9879 \pm 0.0009$, and that of
 the L2 trigger $0.9948 \pm 0.0001$.
 The L3 trigger is not simulated. 
 The L3 trigger efficiency is dominated by differences between the online  and
 off-line reconstruction code efficiency~\footnote{Online algorithms are faster
 but less accurate than the off-line reconstruction code.}.
 The relative L3 efficiency for reconstructing a single muon
 identified by the off-line code
 has been measured to be $0.997 \pm 0.002$~\cite{bishai}.
 The reconstruction efficiencies in the data and in the simulation
  are summarized in Table~\ref{tab:tab2}.
 \begin{table}
 \caption{ Summary of  efficiencies for reconstructing $B^\pm$ candidates in the data
 and the simulation. The last column indicates the corrections applied to the simulated
 acceptance and used to derive  $\cal A_{\rm corr}$ in Table~\ref{tab:tab1}.
} 
\begin{center}
\begin{ruledtabular}
\begin{tabular}{lccc}
  Source   & Data & Simulation & Corr. \\
  COT tracking     & $(0.996 \pm 0.006)^3$ &   $(0.998 \pm 0.002)^3$     & $1.00\pm 0.02$   \\
  Kaon interaction &                       &                             & $1.000 \pm 0.003$\\
  CMU acc. and eff.&  $(0.6251 \pm 0.0047)^2$  & $(0.6439 \pm 0.0004)^2$ & $0.942 \pm 0.014$     \\
  L1 CMU primitives   &  $(0.9276 \pm 0.0005)^2$  & $(0.8369 \pm 0.0004)^2$ & $1.228 \pm 0.002$   \\
  L1  eff.         &  $0.9879 \pm 0.0009$      & 0.9868                  &  $1.0011 \pm 0.0009$  \\
  L2 eff.          & $0.9948 \pm 0.0001$       & 0.9939                  & $1.0009 \pm 0.0001$   \\
  L3 eff.          & $(0.997 \pm 0.002)^2$     & 1                       & $0.994 \pm 0.004$   \\
  Total & $0.328\pm 0.008$                   & $0.283 \pm 0.002$      & $1.152 \pm 0.029$ \\
\end{tabular}
 \end{ruledtabular}
 \end{center}
 \label{tab:tab2}
 \end{table}
 \subsection{Results}
 \label{sec:res}

 The differential cross section $d\sigma /dp_T$ is calculated as
  \begin{equation}
 \frac {d\sigma (B^+)}{dp_T}= \frac{ N/2}{ \Delta p_T \times {\cal L} \times {\cal A_{\rm corr}}  
  \times BR }
\end{equation}
 where $N$ is the number of $B^\pm$ mesons
  determined from the likelihood fit to the invariant mass
 distribution of the $J/\psi \;K^\pm$ candidates in each $p_T$ bin.
  The factor 1/2 accounts for the fact that both $B^+$ and $B^-$ mesons are used
  and assumes
 $C$ invariance at production. The bin width  $\Delta p_T$ and ${\cal A_{\rm corr}}$,
 the geometric and kinematic
 acceptance that includes trigger and tracking efficiencies measured with
 the data, are  listed in Table~\ref{tab:tab1}. The integrated luminosity of the
 data set is $ {\cal L}= 739 \pm 44$ pb$^{-1}$. The branching ratio $BR=
  (5.98 \pm 0.22) \times 10^{-5}$  is derived from
  the branching fractions
  $BR( B^\pm \rightarrow J/\psi \; K^\pm)=(1.008 \pm 0.035) \times 10^{-3} $
  and  $BR( J/\psi \rightarrow \mu^+\;\mu^-)=(5.93 \pm 0.06) \times 10^{-2} $~\cite{pdg}.

  The measured $B^+$ differential cross section as a function
 of its transverse momentum is listed in Table~\ref{tab:tab3} .
 \begin{table}
 \caption{Observed differential cross section, $d\sigma/dp_T$ (nb/$\gevc$),
 for $B^+$ mesons with rapidity $|y| \leq 1$. Errors are the sum in quadrature
  of statistical errors (shown in parentheses) and systematic errors due to
 luminosity (6\%), branching ratios (3.6\%),   acceptance (2.5\%),
  and fitting procedure (2.0\%). Systematic errors 
  are not $p_T$ dependent.
 The integrated cross section for $p_T \geq 25\; \gevc$ is
  $21.7 \pm 3.7$  nb.
} 
\begin{center}
\begin{ruledtabular}
\begin{tabular}{cccc}
  $<p_T>$  ($\gevc$)  & Events & Acceptance (\%) & $d\sigma/dp_T$                    \\
  7.38  &  $ 2792\pm 186$ &   $1.780 \pm 0.045$ &  $591.7 \pm 59.0$ (39.3 stat. )   \\
  10.38 & $2373 \pm 110$ & $4.405 \pm 0.111$    &  $203.2 \pm 17.8$ (9.4 stat.)      \\
  13.39 & $1365 \pm 66$  &  $6.872 \pm 0.173$   &  $74.9 \pm 6.6$ (3.6 stat.)          \\
  19.10 &  $1390\pm 63$  &  $10.16 \pm 0.25$   &   $15.5 \pm 1.3$ (0.7 stat.)          \\
  $\geq 25$      & $277 \pm 44$   & $14.42 \pm 0.36$       &                            \\
\end{tabular}
 \end{ruledtabular}
 \end{center}
 \label{tab:tab3}
 \end{table}
 The  integrated cross section is
 \begin{equation}
 \sigma_{B^+} (p_T \geq 6.0 \,\; \gevc, |y|< 1) = 2.78 \pm 0.24  \; \mu{\rm b},
  \end{equation}
 where the 8.8\%  error is the sum in quadrature of the 6\% error on the integrated luminosity,
  the 3.6\% uncertainty of the $B^+ \rightarrow J/\psi \; K^+$
 and $J/\psi \rightarrow \mu^+\; \mu^-$ branching fractions, 
 the $2.5$\%  uncertainty of the acceptance calculation,
 the 2\% systematic uncertainty of the fit, and
  the $4.4$\% statistical error.

\begin{figure}
 \begin{center}
 \leavevmode
\includegraphics*[width=\textwidth]{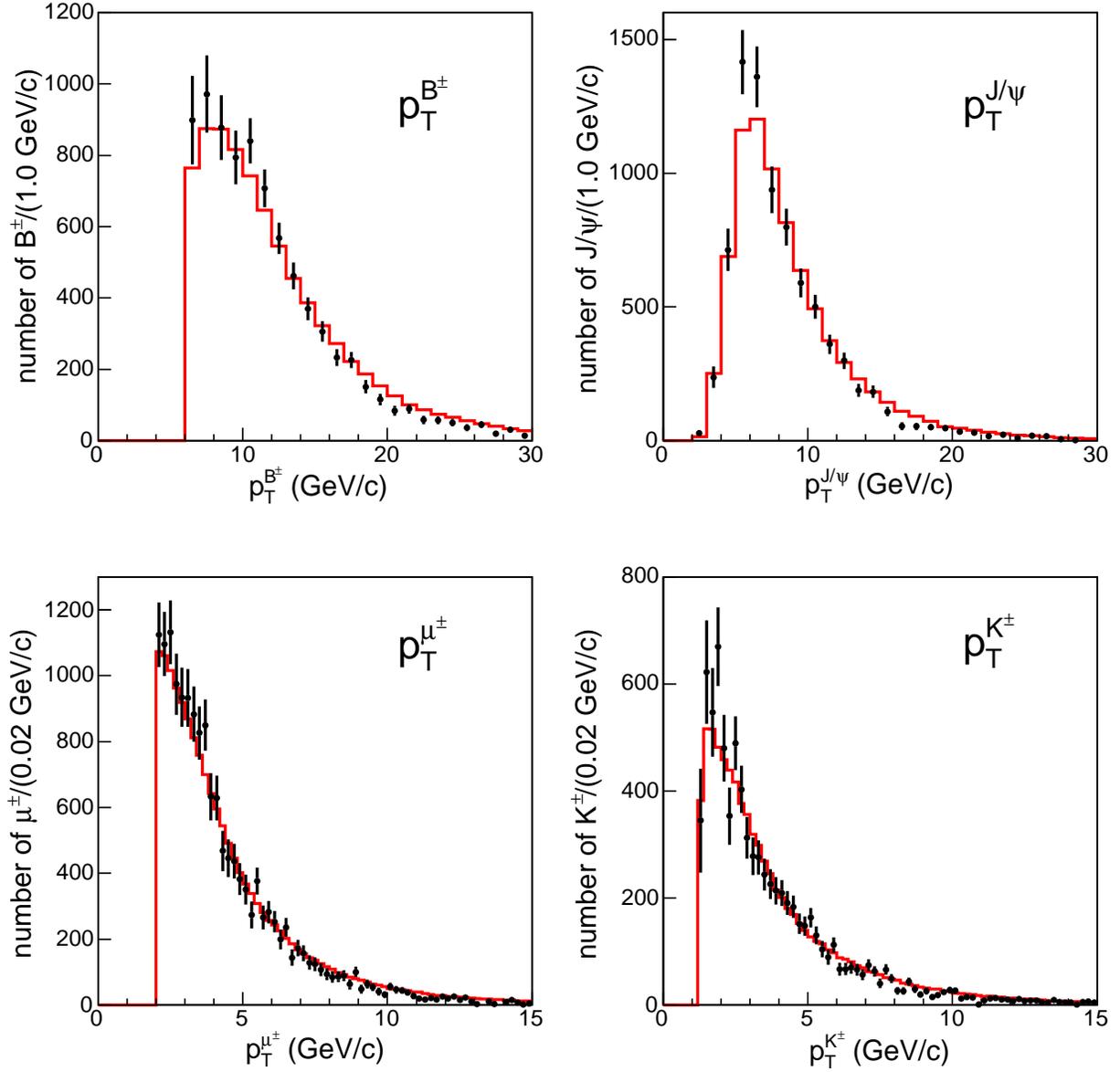}
 \caption[]{Transverse momentum distributions in the data ($\bullet$) and
  simulation (solid histogram). The simulation is normalized to the 
 $B^\pm \rightarrow \mu^+ \; \mu^- \; K^\pm$ signal observed in the data (see text).}
 \label{fig:fig9}
 \end{center}
 \end{figure}

 For completeness, Figure~\ref{fig:fig9} compares transverse momentum
 distributions in the data and in the simulation, based on the NLO QCD
 prediction~$^2$, that has been used to evaluate the detector acceptance.
 Data and
 simulation are  normalized to the
  same number of events. Each distribution is constructed using $J/\psi K^\pm$ candidates 
  with invariant mass
 in the range $5.255-5.315\; \gevcc$ (region~\#1). The background contribution is subtracted using
 candidates in the mass range $5.18-5.24$ and $5.33-5.425\; \gevcc$. 
 The background normalization is 
  the number of events in region~\#1 minus
  the number of $B^\pm$ candidates determined by
 the fit listed in Table~\ref{tab:tab3}.
  One notes the fair agreement between data and untuned QCD prediction~\footnote{   
 The $p_T$ distributions
 of the $B^\pm$ and $J/\psi$ mesons  in the data are
 slightly softer than those 
  of  the simulation;  this difference is
 not relevant for the result of the study because the
  $B^\pm$ kinematic acceptance has been evaluated for each $p_T^B$-bin and
  the calibration of the simulated acceptance using the data
  do not depend on the muon and kaon transverse momenta.}.
 \section{Conclusions} \label{sec:concl}
 We use the exclusive decay $B^\pm \rightarrow J/\psi \; K^\pm $  to measure the
   $B^+$ production cross section  in $p\bar{p}$ collisions at
 $\sqrt{s}=1960$ GeV. The measurement is based on a sample of $8197 \pm 239 $
 $B^\pm$ mesons selected from 739 pb$^{-1}$ of data
 collected with the CDF~II detector at the Fermilab Tevatron collider.
 The  $B^+$ production cross section is measured to be
  $$ \sigma_{B^+} (p_T \geq 6.0 \,\; \gevc, |y|< 1) = (2.78 \pm 0.24)\; \mu{\rm b}.$$
  
\begin{figure}
 \begin{center}
 \leavevmode
\includegraphics*[width=\textwidth]{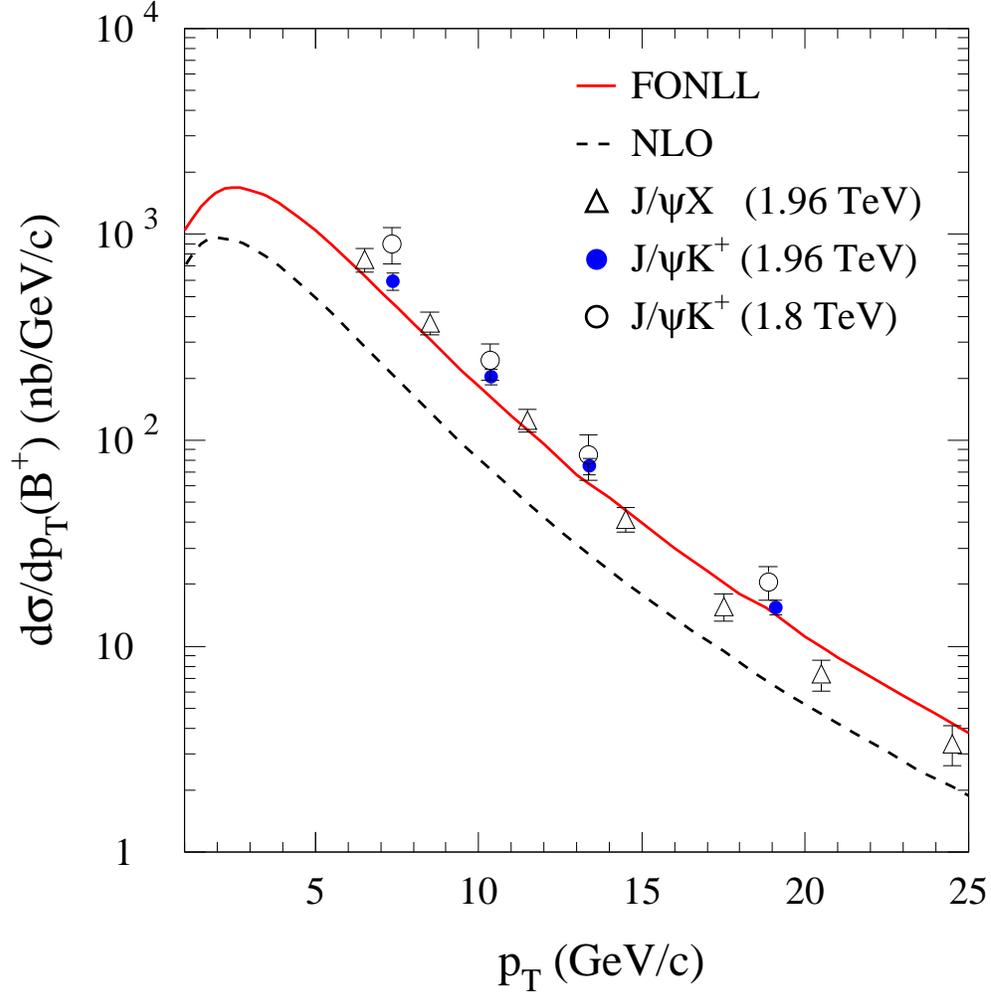}
 \caption[]{Measurements of the  $B^+$ differential cross section ($|y^{B^+}| \leq 1$)
  at the Tevatron
 are compared to the NLO and FONLL theoretical predictions (see text).
 The result of this experiment ($\bullet$) is shown together  with those of
 ($\bigtriangleup$) Ref.~\cite{bishai}
 and ($\circ$) Ref.~\cite{cdfb2}; the result of Ref.~\cite{cdfb2}  has been increased
 by 10\% to account for the expected increase of the cross section from $\sqrt{s}=1.8$ to  
 1.96 TeV~\cite{nlo}.
}
 \label{fig:fig_com}
 \end{center}
 \end{figure}

 To compare with other Tevatron measurements, we
 choose as a theoretical benchmark the NLO QCD prediction~\cite{nlo}
 that uses a $b$-quark mass of
 $m_b=4.75 \; \gevcc$, renormalization and factorization scales
 $\mu_R=\mu_F=\sqrt{p_T^2 +m_b^2}$, the  MRSD$_0$~\cite{mrsd0}
 fit to the parton distribution functions (PDF),  a fragmentation
 fraction $f_u=0.375$, and a fragmentation model
  based on the Peterson fragmentation function with the $\epsilon$ parameter
 set to 0.006. The ratio of the present measurement to this theoretical prediction is
    $2.80 \pm 0.24$. 
 Previous measurements of the single $b$-quark cross section based on the detection of
 $J/\psi$ mesons yield the following ratios
 to the same theoretical prediction:
 $2.9 \pm 0.67$~\cite{cdfb2}, $4.0 \pm 0.6$~\cite{cdfb1},  $ 4.0 \pm 0.4$~\cite{sphi},
 and $3.14 \pm 0.28$~\cite{bishai}.
 In contrast, 
   all  CDF and D${\not\!{\rm O}}$ measurements of the single $b$ production cross
  section that are 
  based upon detection of a lepton from $b$-quark decays~\cite{cdfb3,cdfb4,d0b1,d0b2,d0b3}
 yield a smaller average ratio
  to the same theoretical prediction  (2.2 with a 0.2 RMS deviation~\cite{bstatus}).
  As shown in Fig~\ref{fig:fig_com}, our measurement
  agrees with the value inferred from the
 $J/\psi$ inclusive cross section~\cite{bishai} 
[$  \sigma_{B^+} (p_T \geq 6.0 \,\; \gevc, |y|< 1) = 2.4 \pm 0.4\; \mu{\rm b} $]
  and is within the  range of values predicted by the FONLL QCD
 calculation~\cite{nll,fonll}
  that uses $f_u= 0.389$~\cite{pdg} and the
 CTEQ6M fits to the parton distribution functions~\cite{cteq6m}
 (2.1 $\mu$b with a $\simeq30$\% theoretical uncertainty~\cite{cacc}).
\section{Acknowledgments}
We thank the Fermilab staff and the technical staffs of the participating institutions for their vital contributions. This work was supported by the U.S. Department of Energy and National Science Foundation; 
the Italian Istituto Nazionale di Fisica Nucleare; the Ministry of Education, Culture, Sports,
 Science and Technology of Japan; the Natural Sciences and Engineering Research Council of Canada;
 the National Science Council of the Republic of China; the Swiss National Science Foundation;
 the A.P. Sloan Foundation; the Bundesministerium f\"ur Bildung und Forschung, Germany; 
the Korean Science and Engineering Foundation and the Korean Research Foundation;
 the Particle Physics and Astronomy Research Council and the Royal Society, UK; 
the Institut National de Physique Nucleaire et Physique des Particules/CNRS;
the Russian Foundation for Basic Research; the Comisi\'on Interministerial de Ciencia y Tecnolog\'{\i}a, Spain;
 the European Community's Human Potential Programme under contract HPRN-CT-2002-00292; and the Academy of Finland.

\input{bibliography_bjk.tex}
\clearpage
 \end{document}

%% file: prd_definitions.tex
\def\Z0{${\em Z^0\/}$}

\def\r#1 {$^{#1}$}

\newcommand{\gevc} { {\rm GeV/c}}
\newcommand{\gevcc}{ {\rm GeV/c^2}}

\def\gepsfcentered#1{
  \def\testit{#1}
  \def\lbracket{[}
  \ifx\testit\lbracket
    \let\dofilecmd=\gepsfwithopt
  \else
    \let\dofilecmd=\gepsfnoopt
  \fi
  \dofilecmd}

\def\gepsfnoopt#1{
  \begin{center}
  \leavevmode
  \epsffile{#1}
  \end{center}}

\def\gepsfwithopt#1 #2 #3 #4]#5{
  \begin{center}
  \leavevmode
  \gepsfmaxx=0.94\textwidth
  \epsffile[#1 #2 #3 #4]{#5}
  \end{center}}

\newdimen\gepsfmaxx
\def\epsfsize#1#2{
  \ifnum \epsfxsize=0
    \ifnum \epsfysize=0
      \ifnum #1 > \gepsfmaxx
        \gepsfmaxx
      \else
        #1
      \fi
    \else
      \epsfxsize
    \fi
  \else
    \epsfxsize
  \fi
}

%% file: Oct_2006_Authors_alt_removed.tex
\affiliation{Institute of Physics, Academia Sinica, Taipei, Taiwan 11529, Republic of China} 
\affiliation{Argonne National Laboratory, Argonne, Illinois 60439} 
\affiliation{Institut de Fisica d'Altes Energies, Universitat Autonoma de Barcelona, E-08193, Bellaterra (Barcelona), Spain} 
\affiliation{Baylor University, Waco, Texas  76798} 
\affiliation{Istituto Nazionale di Fisica Nucleare, University of Bologna, I-40127 Bologna, Italy} 
\affiliation{Brandeis University, Waltham, Massachusetts 02254} 
\affiliation{University of California, Davis, Davis, California  95616} 
\affiliation{University of California, Los Angeles, Los Angeles, California  90024} 
\affiliation{University of California, San Diego, La Jolla, California  92093} 
\affiliation{University of California, Santa Barbara, Santa Barbara, California 93106} 
\affiliation{Instituto de Fisica de Cantabria, CSIC-University of Cantabria, 39005 Santander, Spain} 
\affiliation{Carnegie Mellon University, Pittsburgh, PA  15213} 
\affiliation{Enrico Fermi Institute, University of Chicago, Chicago, Illinois 60637} 
\affiliation{Comenius University, 842 48 Bratislava, Slovakia; Institute of Experimental Physics, 040 01 Kosice, Slovakia} 
\affiliation{Joint Institute for Nuclear Research, RU-141980 Dubna, Russia} 
\affiliation{Duke University, Durham, North Carolina  27708} 
\affiliation{Fermi National Accelerator Laboratory, Batavia, Illinois 60510} 
\affiliation{University of Florida, Gainesville, Florida  32611} 
\affiliation{Laboratori Nazionali di Frascati, Istituto Nazionale di Fisica Nucleare, I-00044 Frascati, Italy} 
\affiliation{University of Geneva, CH-1211 Geneva 4, Switzerland} 
\affiliation{Glasgow University, Glasgow G12 8QQ, United Kingdom} 
\affiliation{Harvard University, Cambridge, Massachusetts 02138} 
\affiliation{Division of High Energy Physics, Department of Physics, University of Helsinki and Helsinki Institute of Physics, FIN-00014, Helsinki, Finland} 
\affiliation{University of Illinois, Urbana, Illinois 61801} 
\affiliation{The Johns Hopkins University, Baltimore, Maryland 21218} 
\affiliation{Institut f\"{u}r Experimentelle Kernphysik, Universit\"{a}t Karlsruhe, 76128 Karlsruhe, Germany} 
\affiliation{High Energy Accelerator Research Organization (KEK), Tsukuba, Ibaraki 305, Japan} 
\affiliation{Center for High Energy Physics: Kyungpook National University, Taegu 702-701, Korea; Seoul National University, Seoul 151-742, Korea; and SungKyunKwan University, Suwon 440-746, Korea} 
\affiliation{Ernest Orlando Lawrence Berkeley National Laboratory, Berkeley, California 94720} 
\affiliation{University of Liverpool, Liverpool L69 7ZE, United Kingdom} 
\affiliation{University College London, London WC1E 6BT, United Kingdom} 
\affiliation{Centro de Investigaciones Energeticas Medioambientales y Tecnologicas, E-28040 Madrid, Spain} 
\affiliation{Massachusetts Institute of Technology, Cambridge, Massachusetts  02139} 
\affiliation{Institute of Particle Physics: McGill University, Montr\'{e}al, Canada H3A~2T8; and University of Toronto, Toronto, Canada M5S~1A7} 
\affiliation{University of Michigan, Ann Arbor, Michigan 48109} 
\affiliation{Michigan State University, East Lansing, Michigan  48824} 
\affiliation{Institution for Theoretical and Experimental Physics, ITEP, Moscow 117259, Russia} 
\affiliation{University of New Mexico, Albuquerque, New Mexico 87131} 
\affiliation{Northwestern University, Evanston, Illinois  60208} 
\affiliation{The Ohio State University, Columbus, Ohio  43210} 
\affiliation{Okayama University, Okayama 700-8530, Japan} 
\affiliation{Osaka City University, Osaka 588, Japan} 
\affiliation{University of Oxford, Oxford OX1 3RH, United Kingdom} 
\affiliation{University of Padova, Istituto Nazionale di Fisica Nucleare, Sezione di Padova-Trento, I-35131 Padova, Italy} 
\affiliation{LPNHE, Universite Pierre et Marie Curie/IN2P3-CNRS, UMR7585, Paris, F-75252 France} 
\affiliation{University of Pennsylvania, Philadelphia, Pennsylvania 19104} 
\affiliation{Istituto Nazionale di Fisica Nucleare Pisa, Universities of Pisa, Siena and Scuola Normale Superiore, I-56127 Pisa, Italy} 
\affiliation{University of Pittsburgh, Pittsburgh, Pennsylvania 15260} 
\affiliation{Purdue University, West Lafayette, Indiana 47907} 
\affiliation{University of Rochester, Rochester, New York 14627} 
\affiliation{The Rockefeller University, New York, New York 10021} 
\affiliation{Istituto Nazionale di Fisica Nucleare, Sezione di Roma 1, University of Rome ``La Sapienza," I-00185 Roma, Italy} 
\affiliation{Rutgers University, Piscataway, New Jersey 08855} 
\affiliation{Texas A\&M University, College Station, Texas 77843} 
\affiliation{Istituto Nazionale di Fisica Nucleare, University of Trieste/\ Udine, Italy} 
\affiliation{University of Tsukuba, Tsukuba, Ibaraki 305, Japan} 
\affiliation{Tufts University, Medford, Massachusetts 02155} 
\affiliation{Waseda University, Tokyo 169, Japan} 
\affiliation{Wayne State University, Detroit, Michigan  48201} 
\affiliation{University of Wisconsin, Madison, Wisconsin 53706} 
\affiliation{Yale University, New Haven, Connecticut 06520} 
\author{A.~Abulencia}
\affiliation{University of Illinois, Urbana, Illinois 61801}
\author{J.~Adelman}
\affiliation{Enrico Fermi Institute, University of Chicago, Chicago, Illinois 60637}
\author{T.~Affolder}
\affiliation{University of California, Santa Barbara, Santa Barbara, California 93106}
\author{T.~Akimoto}
\affiliation{University of Tsukuba, Tsukuba, Ibaraki 305, Japan}
\author{M.G.~Albrow}
\affiliation{Fermi National Accelerator Laboratory, Batavia, Illinois 60510}
\author{D.~Ambrose}
\affiliation{Fermi National Accelerator Laboratory, Batavia, Illinois 60510}
\author{S.~Amerio}
\affiliation{University of Padova, Istituto Nazionale di Fisica Nucleare, Sezione di Padova-Trento, I-35131 Padova, Italy}
\author{D.~Amidei}
\affiliation{University of Michigan, Ann Arbor, Michigan 48109}
\author{A.~Anastassov}
\affiliation{Rutgers University, Piscataway, New Jersey 08855}
\author{K.~Anikeev}
\affiliation{Fermi National Accelerator Laboratory, Batavia, Illinois 60510}
% ...
\collaboration{CDF Collaboration\footnote{With visitors from $^a$University of Athens, 
$^b$University of Bristol, 
$^c$University Libre de Bruxelles, 
$^d$Cornell University, 
$^e$University of Cyprus, 
$^f$University of Dublin, 
$^g$University of Edinburgh, 
$^h$University of Heidelberg, 
$^i$Universidad Iberoamericana, 
$^j$University of Manchester, 
$^k$Nagasaki Institute of Applied Science, 
$^l$University de Oviedo, 
$^m$University of London, Queen Mary and Westfield College, 
$^n$Texas Tech University, 
$^o$IFIC(CSIC-Universitat de Valencia), 
}}
\noaffiliation